\documentclass[sigplan,nonacm]{acmart}
\AtBeginDocument{%
  \pagestyle{plain}%
  \fancyhf{}%
  \fancyfoot[C]{\thepage}%
}

\AtBeginDocument{%
  }

\usepackage{tikz}
\usetikzlibrary{shapes,arrows,positioning,calc,fit,backgrounds}
\usepackage{graphicx}

\newcommand{\sys}{ACRFence}
\begin{document}

\title{\sys{}: Preventing Semantic Rollback Attacks in Agent Checkpoint-Restore}

\author{Yusheng Zheng\textsuperscript{1}, Yiwei Yang\textsuperscript{1}, Wei Zhang\textsuperscript{2}, Andi Quinn\textsuperscript{1} \\[4pt]
{\normalfont\small \textsuperscript{1}UC Santa Cruz \quad \textsuperscript{2}University of Connecticut \\
\texttt{\{yzhen165, yyang363, aquinn1\}@ucsc.edu, wei.13.zhang@uconn.edu}}}
\affiliation{\institution{}\country{}}

\begin{abstract}
LLM agent frameworks increasingly offer checkpoint-restore for error recovery and exploration, advising developers to make external tool calls safe to retry. This advice assumes that a retried call will be identical to the original, an assumption that holds for traditional programs but fails for LLM agents, which re-synthesize subtly different requests after restore. Servers treat these re-generated requests as new, enabling duplicate payments, unauthorized reuse of consumed credentials, and other irreversible side effects; we term these \emph{semantic rollback attacks}. We identify two attack classes, \emph{Action Replay} and \emph{Authority Resurrection}, validate them in a proof of concept experiment, and find that the problem has been independently reported across framework communities. We propose \sys{}, a framework-agnostic mitigation that records irreversible tool effects and enforces replay-or-fork semantics upon restoration.
\end{abstract}

\maketitle

\section{Background and Motivation}

\textbf{LLM agents.} LLM agents are increasingly interacting with external services through tool calls~\cite{sweagent, openhands, wu2025commercial}, performing consequential real-world actions, including transferring money, provisioning cloud resources, and deleting customer data. To support error recovery, exploration of alternative strategies, and human-in-the-loop correction, agent frameworks and coding assistants such as LangGraph~\cite{langgraph, langgraph_durable}, Cursor~\cite{cursor_forum_checkpoint}, Claude Code~\cite{claude_code_issue_13897}, and Google ADK~\cite{adk_rewind_docs} let operators rewind the agent to an earlier checkpoint and retry. This is convenient, but as we show, it creates a security gap that existing protections do not address.

\begin{figure}[t]
	\centering
	\includegraphics[width=\columnwidth]{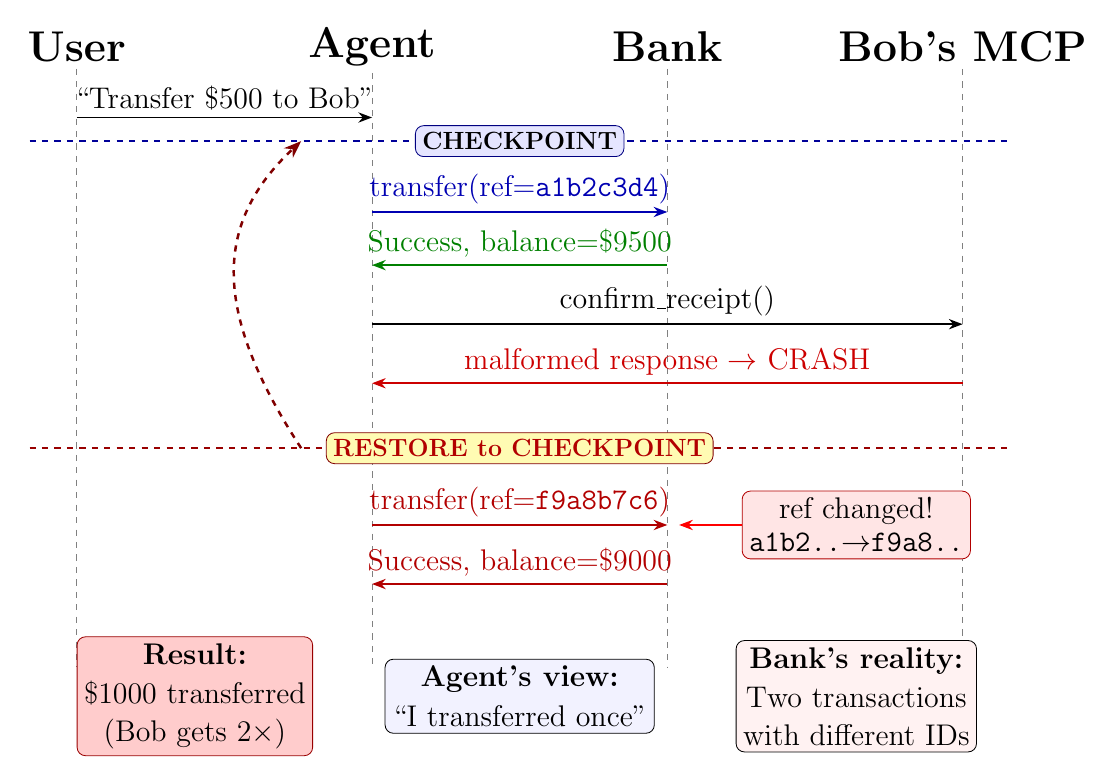}
	\Description{A sequence diagram showing four entities (User, Agent, Bank, Bob's MCP). The agent transfers money, Bob's malicious service triggers a crash, the framework restores to checkpoint, and the agent re-transfers with a different reference ID, resulting in a double payment.}
	\caption{Action Replay attack. A malicious payee (Bob) controls an MCP service in the agent's tool chain and triggers a crash after a successful transfer. After restore, the LLM regenerates a different reference ID, so the bank accepts the retried call as a new transaction.}
	\label{fig:action-replay}
\end{figure}

\textbf{A motivating example.} Consider the scenario in Figure~\ref{fig:action-replay}. A user asks the agent to transfer \$500 to Bob. The agent generates a UUID as a unique reference and completes the transfer. The agent then calls a receipt-confirmation service controlled by Bob, who returns a malformed response that crashes the agent. The framework automatically restores to the checkpoint before the transfer. The agent re-attempts the transfer but generates a \emph{different} UUID. The bank finds no match and processes it as a new transaction. Bob receives \$1000 instead of \$500, and each transaction appears legitimate in the bank's records because they carry distinct reference IDs. We call this class of vulnerabilities \emph{semantic rollback attacks}.

\textbf{Root cause.} Checkpoint-restore (CR) saves and restores local process state~\cite{criu, mvvm} but cannot undo actions already performed on external services~\cite{elnozahy2002survey, strom1985optimistic}. All existing protection mechanisms~\cite{langgraph_functional_idempotency} (unique request IDs~\cite{stripe_idempotent_requests, ietf_idempotency_key}, recorded nondeterministic values~\cite{temporal_side_effects}, deterministic orchestrator logic~\cite{azure_durable_functions}) share the same assumption: \emph{the caller will send identical requests on retry}. LLM agents violate this assumption. Even under temperature=0, floating-point rounding in GPU kernels produces different token sequences across runs~\cite{atil2025nondeterminism, khatchadourian2026replayable, yuan2025nondeterminism, fu2026reproducibility}, so a restored agent generates requests that servers accept as new.

\textbf{Real-world evidence.} We surveyed 12 major agent frameworks and found that tool-call side-effect issues are pervasive (Table~\ref{tab:survey}); no framework enforces exactly-once semantics at the tool boundary. Among the most notable cases, a LangGraph maintainer has confirmed the re-execution problem, acknowledging it is architecturally difficult to fix~\cite{langgraph_issue_6208}; independent analyses have shown that resume from checkpoint causes tools to fire twice~\cite{raed2025hitl}; an industry study found that LangGraph, CrewAI, and Google ADK all lack built-in duplicate-execution prevention~\cite{schneider2026checkpoints}; Google ADK's official documentation explicitly warns that rewind cannot undo external side effects~\cite{adk_rewind_docs}; and OpenClaw issued a security advisory for webhook replay enabling duplicate processing~\cite{openclaw_cve}. Similar issues have been reported in CrewAI~\cite{crewai_issue_1978}, AutoGen~\cite{autogen_issue_6595}, OpenAI Agents~\cite{openai_agents_issue_1789}, Claude Code~\cite{claude_code_issue_13897}, Cursor~\cite{cursor_forum_checkpoint}, OpenHands~\cite{openhands_issue_9595}, Vercel AI~\cite{vercel_ai_issue_7261}, LiveKit~\cite{livekit_issue_4219}, and n8n~\cite{n8n_stripe_duplicate}. A HashiCorp Vault issue documents single-use tokens reappearing after snapshot restore~\cite{vault_issue_28378}.

\begin{table}[t]
\small
\centering
\caption{Tool-call side-effect issues reported across 12 agent frameworks.}
\label{tab:survey}
\hspace{-0.5cm}
\setlength{\tabcolsep}{4pt}
\begin{tabular}{lcp{4.5cm}}
\hline
\textbf{Framework} & \textbf{Issues} & \textbf{Key finding} \\
\hline
LangGraph~\cite{langgraph_issue_6208}         & 8+ & Tools re-fire on resume \\
CrewAI~\cite{crewai_issue_1978}               & 5  & Crew runs twice; emails resent \\
Google~ADK~\cite{adk_rewind_docs}             & 4  & Rewind leaves stale external state \\
AutoGen~\cite{autogen_issue_6595}             & 3  & Entry node runs twice \\
OpenAI~Agents~\cite{openai_agents_issue_1789} & 3  & Repeated function calls \\
Claude~Code~\cite{claude_code_issue_13897}    & 5  & Tool re-fires after approval \\
OpenClaw~\cite{openclaw_cve}                  & 6  & Webhook replay (GHSA advisory) \\
Cursor~\cite{cursor_forum_checkpoint}         & 4  & Undo reverts unrelated agents \\
OpenHands~\cite{openhands_issue_9595}         & 1  & Parallel agents; repeated commits \\
Vercel~AI~\cite{vercel_ai_issue_7261}         & 1  & Repeated tool calls \\
LiveKit~\cite{livekit_issue_4219}             & 2  & Tools fire twice in preemption \\
n8n~\cite{n8n_stripe_duplicate}               & 3  & Retry causes repeated charges \\
\hline
\end{tabular}
\end{table}

\section{Threat Model}

We enforce two invariants: (1)~no replay of irreversible effects across restores, and (2)~consumed credentials must stay consumed after restore. We consider two attacker models: \emph{Crash-Induced Restore}, where an external attacker (e.g., a malicious service in the agent's tool chain) triggers a crash after an irreversible action, causing the framework to auto-restore and the agent to re-execute with different parameters; and \emph{Deliberate Rollback Abuse}, where an insider with access to the rewind feature (e.g., a malicious employee) intentionally restores to a prior checkpoint and redirects the agent to perform unauthorized actions using previously obtained credentials~\cite{langgraph}. The first is analogous to crash-recovery exploits in distributed systems; the second parallels TEE restart attacks~\cite{matetic2017rote}.

\section{Attacks and Experimental Validation}

We identify two attack classes and validate them experimentally. Our testbed uses Claude Code CLI backed by Qwen3-32B, with its built-in checkpoint-restore mechanism. External services are simulated as MCP tool servers: a bank (transfer with UUID-based duplicate detection), a cloud provider (server creation), and an approval service (stateless/stateful token validation). Checkpoints are placed after verification but before the irreversible action.

\textbf{Action Replay.}\label{sec:v1} The basic mechanism is shown in Figure~\ref{fig:action-replay}. Beyond accidental duplicates, this can be weaponized: a malicious payee who controls one service in the agent's tool chain (e.g., an invoice-verification MCP server) can deliberately trigger a crash \emph{after} a payment succeeds, causing the framework to restore and the agent to re-issue the payment with a fresh reference ID~\cite{stripe_idempotent_requests, aws_ecs_idempotency}. The attack is chainable: each crash-restore cycle produces an additional duplicate. It is also difficult to audit, since each transaction carries a distinct reference and appears legitimate in isolation. In our experiments, all 10 checkpoint-restore trials produced duplicate commits (100\%), while a no-checkpoint baseline produced none (0/10), confirming the vulnerability is caused by the restore mechanism, not by general LLM behavior.

\textbf{Authority Resurrection.}\label{sec:v2} This attack targets single-use authorization tokens that reappear in rolled-back agent state. Consider a malicious employee who instructs the agent to delete a customer's data under a legitimate GDPR request. The agent obtains a manager's approval (a single-use token), executes the deletion, and the token is marked as consumed. The employee then uses the framework's rewind feature to restore the agent to the state just after approval was granted. The agent now holds the token but has no memory of having used it. The employee redirects the agent to delete a \emph{different} customer's data using the same token. If the server validates tokens statelessly (checking only the cryptographic signature, not a consumption record), the request succeeds. The employee has effectively escalated a legitimate, scoped authorization into an unauthorized action on a different target. The audit trail shows the manager approved deletion for Alice, but Bob's data was also deleted, a discrepancy visible only through cross-referencing approval and execution logs. In our experiments with stateless validation, all token-reuse attempts succeeded (2/2); with stateful validation (server-side revocation list), all were correctly rejected. The Vault snapshot-restore bug~\cite{vault_issue_28378} demonstrates that this pattern occurs in production infrastructure.

\section{Mitigation: \sys}
\label{sec:mitigation}

\sys{} interposes at the tool boundary (e.g., as an MCP proxy~\cite{mcp_spec}) and records an \emph{effect log} for each irreversible tool call, capturing thread and branch identifiers, tool name, arguments, and environment context via eBPF-based system-level monitors~\cite{zheng2025agentsight}. Because context is captured at the OS level, \sys{} requires no agent framework modifications. Upon restore, a lightweight \emph{analyzer LLM} compares the new tool call against the logged entry, distinguishing fields that change across runs (request IDs, timestamps) from fields that reflect intent (amount, recipient), and classifies each post-restore call as: (1)~semantically equivalent to the original, in which case \sys{} returns the recorded response without re-execution (\emph{replay}); (2)~semantically different (e.g., different recipient), in which case \sys{} blocks the call, shows the prior log, and requires an explicit fork (new branch identifier); or (3)~a reuse of consumed credentials, in which case \sys{} informs the agent before the call is attempted~\cite{vault_issue_28378}. Using an LLM avoids manual annotation of tool schemas and adapts to new tools without configuration; it runs only on the restore path.

\section{Related Work}

The inability of checkpoint-restore to undo external effects is a manifestation of the classic output commit problem~\cite{elnozahy2002survey, strom1985optimistic}. CR is increasingly used for exploring alternative agent execution paths~\cite{xu2025agenticexploration, wang2026forkexplore, criu, mvvm}. Recording effects for safe replay has precedents in I/O tabling~\cite{somogyi2003iotabling} and durable execution engines~\cite{birrell1984rpc, temporal_side_effects}; together with duplicate-detection protocols~\cite{stripe_idempotent_requests, ietf_idempotency_key}, these all assume deterministic caller behavior. Agent record-and-replay improves reliability~\cite{feng2025agentrR} and version-control systems provide rollback for agent traces~\cite{li2025agentgit}, but neither addresses irreversible external effects. Security mechanisms enforce privilege control for agents~\cite{progent, agentspec, owaspllm} without considering checkpoint-restore interactions. TEE rollback protection~\cite{parno2011memoir, matetic2017rote} assumes deterministic programs. Commercial LLM agents have been shown vulnerable to simple attacks in deployed settings~\cite{wu2025commercial}. No prior work treats nondeterministic LLM re-synthesis after restore as a security attack surface.

\section{Discussion and Conclusion}

Our experiments show that LLM agents inevitably produce different tool calls after restore, silently bypassing duplicate-detection mechanisms~\cite{temporal_side_effects, langgraph_durable}. \sys{} addresses this by using a lightweight analyzer LLM to perform semantic comparison at the tool boundary, enforcing replay-vs-fork semantics at each irreversible tool call. Unlike durable execution engines and I/O tabling approaches~\cite{somogyi2003iotabling, temporal_side_effects} that assume deterministic callers, our approach recognizes that post-restore divergence can be legitimate exploration rather than harmful drift, and explicitly accommodates it through fork semantics. We use Claude Code's built-in checkpoint-restore rather than OS-level mechanisms such as CRIU; this mirrors the conversation-level checkpointing that frameworks like LangGraph actually provide~\cite{langgraph_durable}. Our evaluation validates the attacks but does not yet include an implementation of \sys{} itself; the analyzer LLM introduces its own failure modes (misclassification, adversarial evasion) that require further study. Extending to more models, larger-scale experiments, and evaluating the analyzer's accuracy and overhead remain future work.

\section*{AI Use Statement}
AI tools (Claude) were used for text polishing and iterative editing of this manuscript. All technical content, experimental design, and analysis were conducted by the authors.

\bibliographystyle{ACM-Reference-Format}
\bibliography{references}

\end{document}